\documentclass[review]{elsarticle}
\usepackage{a4wide}
\usepackage{natbib} 			
\usepackage{graphicx}
\usepackage{amsfonts}
\usepackage{amsmath}
\usepackage{booktabs}
\usepackage{epsfig}
\usepackage[colorlinks]{hyperref}
\hypersetup{citecolor=blue,linkcolor= blue}

\begin{document}
\title{
How are rescaled range analyses affected by different memory and distributional properties? A Monte Carlo study}
\author{Ladislav Kristoufek}
\ead{kristoufek@ies-prague.org}
\address{Institute of Economic Studies, Charles University, Opletalova 26, 110 00, Prague, Czech Republic\\
Institute of Information Theory and Automation, Academy of Sciences of the Czech Republic, Pod Vodarenskou Vezi 4, Prague 8, 182 08}

\begin{abstract}
In this paper, we present the results of Monte Carlo simulations for two popular techniques of long-range correlations detection -- classical and modified rescaled range analyses. A focus is put on an effect of different distributional properties on an ability of the methods to efficiently distinguish between short and long-term memory. To do so, we analyze the behavior of the estimators for independent, short-range dependent, and long-range dependent processes with innovations from 8 different distributions. We find that apart from a combination of very high levels of kurtosis and skewness, both estimators are quite robust to distributional properties. Importantly, we show that R/S is biased upwards (yet not strongly) for short-range dependent processes, while M-R/S is strongly biased downwards for long-range dependent processes regardless of the distribution of innovations.\\
\end{abstract}


\maketitle


\section{Introduction}
Examination of long-range dependence has become more frequent in various fields in recent years. The portfolio of disciplines has grown wide and varies across finance \citep{DiMatteo2007,Alvarez-Ramirez2008,Kristoufek2010a}, cardiology \citep{Yeh2009,Shiogai2010}, vascular science \citep{Liao2010}, DNA sequences \citep{Jose2005}, pollution examination \citep{Windsor2001}, river flow fluctuations \citep{Movahed2008}, geochemistry \citep{Zuo2009}, earthquakes waiting times \citep{Chen2008}, climatology \citep{Rehman2009} and geomagnetism \citep{Hayakawa2004}. Hurst exponent $H$, $0<H<1$, is a measure of long-range dependence for a stationary process with an autocorrelation function $\rho(k)$ decaying as $\rho(k) \propto k^{2H-2}$ with a lag $k\rightarrow \infty$. $H=0.5$ indicates two possible processes -- either an uncorrelated or a short-range dependent process. Autocovariances of the uncorrelated process are insignificant for all non-zero lags whereas for the short-range dependent process (e.g. ARIMA processes, for details, see \cite{Box1970}), autocovariances are significant at low lags and insignificant at high lags and decay exponentially.  If $H > 0.5$, the process has positive correlations at all lags and is said to be long-range dependent with positive correlations or persistent. On the other hand, if $H < 0.5$, the process is long-range dependent with negative correlations or anti-persistent. Autocovariances of long-range dependent processes are hyperbolically decaying and are either non-summable for the persistent or summable for the anti-persistent processes \citep{Beran1994,Lillo2004,Embrechts2002,Mandelbrot1968,Barkoulas2000}.  

In a majority of applied research papers focusing on long-range dependence in a time series, the results are interpreted on a basis of simple comparison of the estimated Hurst exponent $H$ with the value of $0.5$. If the estimate deviates from 0.5, the studied process is claimed to be long-range dependent. However, the estimators of Hurst exponent are usually either biased or have high variance for finite samples as shown in several studies (e.g. \cite{Weron2002,Couillard2005,Grech2005,Lo1991,Barunik2010,Kristoufek2010}). The estimators can also be biased by a presence of short-range dependence in an underlying process \citep{Morariu2007,Lo1991}. However, distinguishing between the short and the long-range dependence is crucial for time series analysis as each type of dependence implies different features of the series (see \cite{Beran1994} for discussion). To add to the discussion, we test two methods which are closely related and one of them was constructed to be robust against the short-range dependence -- the rescaled range analysis of Hurst \cite{Hurst1951}, which is also called the "classical", and the modified rescaled range analysis of Lo \cite{Lo1991}. Finite sample properties of the two methods and efficiency of the distinguishing between the types of dependence have been partially discussed by Teverosky \textit{et al.} \cite{Teverosky1999}. However, the simulated processes always have normally distributed innovations. Moreover, Teverosky \textit{et al.} \cite{Teverosky1999} tested the ability of the modified rescaled range analysis to adjust the rescaled ranges from the classical form of the method rather than its ability to estimate Hurst exponent $H$. To fill the gap, we test the robustness of the rescaled range analyses for independent, short-range dependent and long-range dependent processes with innovations from 8 different distributions -- standard normal, log-normal, Cauchy, log-$t$, gamma, inverse gamma, Laplace and log-Laplace. By a choice of distribution, we discuss on the robustness of both methods to skewness, kurtosis and infinite moments as well as to different types of memory.

The paper is structured as follows. In Section 2, we present and describe both techniques in detail. In Section 3, we present important parameters of Monte Carlo simulations methodology and describe the chosen distributions. In Section 4, we show results of the Monte Carlo simulations for time series lengths from 512 to 16,384 observations for both tested methods and different types of distributions of the underlying process. We find four important outcomes -- extreme excess kurtosis combined with extreme skewness causes the estimates to be downward biased; M-R/S in its simplest way is not able to distinguish between short and long memory efficiently; R/S is biased by the short-range dependence in the series but in a lesser way than standardly claimed; and apart from very extreme cases of high skewness and leptokurtosis, both methods are quite robust to different distributional properties. Section 5 concludes.

\section{Classical and modified rescaled range analysis}
The classical rescaled range analysis (R/S) was developed by Edwin Hurst while working as an engineer in Egypt \citep{Hurst1951} and was later introduced to the financial time series by Beno\^{i}t Mandelbrot \citep{Mandelbrot1968,Mandelbrot1968a,Mandelbrot1970}. For time series $X_t$ of length $T$ with $t=0,1,\ldots,T$, Hurst exponent $H$ can be estimated from behavior of rescaled ranges of the series. In the procedure, one takes the increments of the time series $x_t=X_t-X_{t-1}$ with $t=1,\ldots,T$ and divides them into $N$ adjacent sub-periods $I_n$, for $n=1,\ldots,N$, of length $\upsilon$ while $N\upsilon = T$. For each sub-period, one calculates an average value and constructs a new series of accumulated deviations from the arithmetic mean values -- a profile $y_t$. The procedure follows in a calculation of the range of the profile $R_{I_n}=max_{I_n}(y_t)-min_{I_n}(y_t)$ and a standard deviation $S_{I_n}$ of the original increment time series for each sub-period. Each range is standardized by the corresponding standard deviation and forms the rescaled range so that the average rescaled range for the given sub-period length $(R/S)_\upsilon$ is calculated \citep{Peters1994}. Rescaled ranges scale as $(R/S)_\upsilon \propto \upsilon^H$.

R/S was shown to be sensitive to heteroskedasticity and short-range dependence in the underlying process \citep{Lo1999,Alfi2008}. To deal with the problem, Lo \cite{Lo1991} proposed the modified rescaled range analysis (M-R/S). The procedure differs in the definition of the standard deviation which deals with both heteroskedasticity and short-range dependence. The modified standard deviation is defined with a use of auto-covariance $\gamma_j$ of the selected sub-interval $I_n$ for lag $j=1,\ldots,\xi$ as

\begin{equation}
\label{eq1}
S^M_{I_n}=\sqrt{S^2_{I_n}+2\sum_{j=1}^\xi\gamma_j\left(1-\frac{j}{\xi+1}\right)}.
\end{equation}

Thus, R/S turns into a special case of M-R/S with $\xi=0$. The most problematic and also the crucial issue of the new standard deviation measure is the number of lags $\xi$ which are used for its estimation. On one hand, if the chosen lag is too low, it omits the lags which may be significant and therefore the estimates of the rescaled ranges and Hurst exponent are still biased. On the other hand, if the used lag is too high, the estimates of the rescaled range differ significantly from the true values \citep{Teverosky1999,Wang2006}.

One can either estimate the rescaled ranges and Hurst exponents for arbitrary values of $\xi$ \citep{Zhuang2000,Alptekin2006} or use an automatic estimator of $\xi$ for each sub-interval of the length $\upsilon$. Lo \cite{Lo1991} suggests to set the optimal lag $\xi^{\ast}$ based on the sample first-order autocorrelation coefficient $\widehat{\rho(1)}$ of the sub-interval $I_n$ as (where $\lfloor\rfloor$ is the nearest lower integer operator) 

\begin{equation}
\label{eq2}
\xi^{\ast}=\left\lfloor\left(\frac{3\upsilon}{2}\right)^{\frac{1}{3}}\left(\frac{2\widehat{\rho(1)}}{1-\widehat{\rho(1)}^2}\right)^{\frac{2}{3}}\right\rfloor.
\end{equation}

Of course, Eq. \ref{eq2} assumes that the underlying process is some specification including AR(1) process. Teverosky \textit{et al.} \cite{Teverosky1999} showed that M-R/S is biased towards rejecting long-range dependence in the process when the high number of lags is used. The bias has been shown on the behavior of the estimated rescaled ranges. In our simulations, we examine whether such a bias is also present for the estimation of Hurst exponent $H$.

Finite sample properties of the rescaled range analysis and the modified rescaled range analysis were discussed in several papers. Most importantly, the condition for the process to be characterized as uncorrelated, which is $H = 0.5$, was shown to hold only asymptotically. For finite samples, the estimated Hurst exponent can differ significantly. Anis \& Lloyd \cite{Anis1976} showed that for the finite samples, expected value of the rescaled range behaves as 
\begin{equation}
\label{eq_AL}
E(R/S)_{\upsilon}=\frac{\Gamma(\frac{\upsilon-1}{2})}{\sqrt{\pi}\Gamma(\frac{\upsilon}{2})}\sum_{i=1}^{\upsilon-1}\sqrt{\frac{\upsilon-1}{i}}.
\end{equation}

Further, Peters \cite{Peters1994} proposed to adjust the expected value of the rescaled range, which was claimed to be better for low scales $\upsilon<50$, to $$E(R/S)_{\upsilon}=\frac{\upsilon-{1/2}}{\upsilon}\sqrt{\frac{2}{\upsilon\pi}}\sum_{i=1}^{\upsilon-1}\sqrt{\frac{\upsilon-1}{i}}.$$

However, Couillard \& Davison \cite{Couillard2005} and Grech \& Mazur \cite{Grech2005} showed that the expected values based on the Monte Carlo simulations are closer to the method of Anis \& Lloyd \cite{Anis1976}. Importantly, they also showed that the standard deviation of the estimates does not converge according to the central limit theorem, which is as a square root of the time series length. Such finding is essential for the confidence interval construction as the slower convergence of standard deviation makes the confidence intervals wider.
   
\section{Monte Carlo simulations methodology}
In this section, we briefly describe the types of simulated processes with different kinds of memory, depict types of distributions of the innovations $\varepsilon_t$ for the simulated processes -- standard normal, log-normal, Cauchy, log-$t$, gamma, inverse gamma, Laplace and log-Laplace -- and specify the crucial parameters of Monte Carlo simulations.  

\subsection{Processes}

To uncover whether R/S and M-R/S are able to efficiently distinguish between short and long-range memory, we simulate three different types of processes -- independent, short-range dependent and long-range dependent. For independent process, we simply choose $i.i.d.$ process, i.e.  $x_t=\varepsilon_t$. For short-range dependent process, we use an autoregressive process with $AR(1)$ specification, i.e. $x_t=\theta x_{t-1}+\varepsilon_t$. And for long-range dependent process, we simulate fractionally integrated ARMA process ARFIMA($0,d,0$) defined as $x_{t}=\sum_{i=1}^{\infty}{a_i(d)x_{t-i}+\varepsilon_t}$ where $0<d<0.5$ is a free parameter, related to Hurst exponent as $H=d+0.5$, and $a_i(d)=d\Gamma(i-d)/(\Gamma(1-d)\Gamma(1+i))$. Infinity is trimmed to 100 lags in the simulations. Note again that the noise terms $\varepsilon_t$ are drawn from various distributions described in the following subsection. Process $X_t$ is then taken as an integrated series $X_t=\sum_{i=1}^{t}{x_i}$ for $t=1,\ldots,T$ and $X_0=0$.

\subsection{Distributions description}

Processes based on normal distribution are used in a majority of models for simplicity. We use $N(0,1)$ as a benchmark and for a comparison as well since majority of sampling properties studies base the results on such process.

Log-normal distribution has several properties which are in hand with stylized facts such as the fat tails and the skewness of e.g. financial returns \citep{Cont2001}. For $X\sim N(0,1)$, $Y=e^{X}$ is log-normally distributed. Moreover, we subtract a unity from $Y$ so that the distribution has zero mean and a minimum of $-1$. We apply the subtraction to all distributions to have the mean of zero.

Cauchy distribution is a stable distribution with $\alpha=1$. For our purposes, we use the standard Cauchy distribution, which is equal to the Student's $t$ distribution with one degree of freedom, defined on the basis of the stable distributions as $S(1,0,\sqrt{2}/2,0)$. As Cauchy distribution is symmetric, we isolate the effect of the heavy tails from the one of the skewness.

Log-$t$ distribution has similar properties to log-normal distribution. However, log-$t$ distribution is more skewed and has heavier tails than log-normal distribution. If $X\sim t_{n}$ is Student's $t$ distributed with $n$ degrees of freedom, then $Y=e^{X}$ is log-$t$ distributed with $n$ degrees of freedom. In simulations, we use $n=5$ to have the final distribution with finite mean, variance, skewness and kurtosis.

Gamma distribution is described by parameters $k$ and $\theta$ and can be seen as a sum of $k$ independent exponential distributions with mean $\theta$ for $k\in\mathbb{N}$. Such gamma distribution has a mean of $k\theta$, a standard deviation of $\sqrt{k\theta^2}$, a skewness of $2/\sqrt{k}$, an excess kurtosis of $6/k$  and is noted as $G(k\theta,\sqrt{k\theta^2})$.  For the simulations, we set $k=4$ and $\theta=0.25$ and subtract a unity to obtain $G(0,2)$ with the minimum of $-1$, the skewness of 1 and the excess kurtosis of 3/2.

Inverse gamma distribution is obtained from an inverse of gamma distribution. If $X\sim G(k\theta,\sqrt{k\theta^2})$, then $Y=1/X$ is inverse gamma distributed with a mean $\theta/(k-1)$, a standard deviation of $\sqrt{\theta^2/((k-1)^2(k-2))}$, a skewness of $4\sqrt{k-2}/(k-3)$ and an excess kurtosis of $(30k-66)/(k-3)(k-4)$. We set the parameters as for the gamma distribution and subtract unity to obtain skewed series with an undefined kurtosis.

Laplace distribution is similar to normal distribution but is based on absolute deviations from a mean rather than squared deviations. Laplace distribution is sometimes called double-exponential as it is formed from two exponential distributions which equal at and are symmetric around the mean. The distribution is then symmetric and leptokurtic with excess kurtosis of 3. We use the standard Laplace distribution with zero mean and a standard deviation of one -- $L(0,1)$.

Log-Laplace distribution has again similar properties to the other ones based on exponential of the symmetric distributions -- a nonzero skewness and an excess kurtosis. If $X\sim L(0,1)$ is Laplace distributed, then $Y=e^{X}$ is log-Laplace distributed. As for the other exponential transformations, we subtract a unity to arrive at zero mean.

\subsection{Choice of parameters}
The choice of parameters is crucial for the final results of the simulations and the estimates. The sub-length $\upsilon$ is set equal to the power of a fixed integer value \citep{Weron2002}. We use a basis $b=2$, a minimum power $p_{min}$ and a maximum power $p_{max}$ so that we get $\upsilon = b^{p_{min}},\ldots,b^{p_{max}}$, where $b^{p_{min}}\ge b$ is a minimum scale and $b^{p_{max}}\le T$ is a maximum scale. We use the minimum scale $\upsilon_{min}=2^5$ and the maximum scale $\upsilon_{max}=T$. Such choice avoids biases of the low scales where the standard deviations can be estimated inefficiently \citep{Peters1994,Grech2005,Matos2008}
. The used intervals are contingent and non-overlapping (see \cite{Ellis2007} for discussion).
 
In the following section, we simulate 1,000 time series with lengths ranging from $2^9$ to $2^{14}$ for the standard normal distribution and estimate Hurst exponent based on R/S and M-R/S for each series. We apply the same procedure for log-normal, Cauchy, log-$t$, gamma, inverse gamma, Laplace and log-Laplace distributions. For each distribution, we construct three different types of processes -- independent, ARFIMA with $d=0.25$ and AR(1) with $\theta=0.25$ -- and comment on an accuracy of the methods for uncorrelated, short-term correlated and long-term correlated processes. To compare the quality of estimators for different types of processes and distribution of innovations, we compare their bias $\theta=\langle\widehat{H_i}-E(H)\rangle$, variance $\sigma^2_H=\langle\widehat{H_i^2}\rangle-\langle\widehat{H_i}\rangle^2$ and mean squared error $MSE=\langle(\widehat{H_i}-E(H))^2\rangle=\theta^2+\sigma_H^2$, where $\widehat{H_i}$ is the estimated Hurst exponent for the simulated series, $E(.)$ stands for the expected value, and $\langle \rangle$ is an average operator for $i=1,\ldots,1000$. Note that the expected value of $H$ for independent and short-range dependent processes is not 0.5 but the value needs to be calculated from Eq. \ref{eq_AL}. For long-range dependent processes, we expect $H=0.75$ for ARFIMA(0,0.25,0). These expected values are used to calculate bias and MSE in the following section.

\section{Monte Carlo simulations results}
The results of Monte Carlo simulations are summarized in Tables \ref{tab1} - \ref{tab6}. We now discuss the results for the three types of processes separately. For $i.i.d.$ processes, we expect that the methods yield same results. For long-range dependent processes, the methods should give similar results. However, M-R/S is expected to underestimate the true Hurst exponent because long-range dependence can be misinterpreted by the method as a strong short-range dependence. Such underestimation should be only small to keep the method usable. For short memory processes, R/S is expected to be biased upwards while M-R/S is expected to be unbiased.

\subsection{Independent process}

Tables \ref{tab1} and \ref{tab2} contain the results for $i.i.d.$ simulations for R/S and M-R/S, respectively. Regarding the bias, there are three interesting findings. First, R/S tends to overestimate the Hurst exponent, while M-R/S method underestimates it. However, the biases are of an order of $10^{-3}$ and thus rather negligible. Second, the processes based on logarithmic distributions lead to stronger underestimation of the Hurst exponents. This is true for log-$t$ and log-Laplace distributions but not for log-normal distribution for both R/S and M-R/S. As the distributions have both non-zero skewness and excess kurtosis, the bias is an implication of one or both of theses. Similar yet not so strong bias is observed for Cauchy distribution. As Cauchy distribution has infinite variance (and is symmetric even though it has an undefined skewness), the downward bias seems to be caused by a combination of excess kurtosis and strong skewness. However, the values of skewness and kurtosis have to be rather high to cause such a bias since other distributions (log-normal, Laplace, gamma and inverse-gamma) are not affected in such a way. Third, the bias in general does not show any tendency to decrease with an increasing time series length.

As the biases are rather small, the qualitative results for variance and MSE of the estimators practically overlay. Here, we find two important outcomes. First, the estimators for log-$t$ and log-Laplace distributions show the smallest variance and MSE. Second, the variance and MSE of the estimators decrease with an increasing time series length for all distributions and both rescaled range methods. Overall, the two methods do not differ significantly for $i.i.d.$ processes analyzed here.

\subsection{ARFIMA($0,0.25,0$)}

The results are more interesting for the ARFIMA processes (Tables \ref{tab3} and \ref{tab4}). For both R/S and M-R/S, the bias of estimators increases (in absolute terms) with a time series length. For example for R/S and a standard normal distribution, the estimator is unbiased for a length of $2^9$ but is strongly biased for $2^{14}$ with a bias of -0.06. This finding does not change for any distribution used. Such a result is most likely caused by the fact that the rescaled range analysis does not have an expected value of 0.75 for ARFIMA process with $d=0.25$. Unfortunately, these expected values are not yet analytically solved. As for the case of independent processes, when the expected values are derived from Eq. \ref{eq_AL}, the expected value of Hurst exponent decreases with increasing time series length. This is reflected in the increasing bias of the estimators for ARFIMA simulations.

More importantly, the bias differs remarkably for R/S and M-R/S. For the standard normal distribution with the shortest length, the bias is 100 times higher for M-R/S than for R/S. Even though this is a rather extreme case majorly caused by the fact that R/S is almost unbiased for the normal distribution, the differences are quite large even for the other distributions. R/S estimator is most biased for log-$t$ distribution, yet M-R/S estimator is more than 5 times more biased. For the others, the bias of M-R/S is around 20 times the bias of R/S. This huge disproportion decreases with an increasing time series length. Nevertheless, the M-R/S bias is between 2-3 times higher than the one of R/S even for the length of $2^{14}$ for all distributions.

Moving to the absolute terms, the bias of R/S is insignificant for the shortest time series for all distributions and increases to around -0.06 for the longest time series for all but the log-$t$ distribution with a bias of -0.0969. For M-R/S, the bias is smallest for the shortest series (around -0.1) and increases to approximately -0.15 for the series of $2^{14}$ observations for all examined distributions. This is a very important finding which indicates that M-R/S misinterprets the long-memory of the process as the short-memory. Even though the method is constructed to rid of the strong short-term correlations, our results show that the correction for the short memory is way too strong. This practically disqualifies the method from being used for long memory analysis.

The variances of the M-R/S estimator are lower than the ones of the R/S for all distributions and all time series lengths (the R/S variances are around 0.5 times higher than these of M-R/S). The variance decreases with an increasing series length for both methods and all distributions. Again, the variances for log-$t$ and log-Laplace distributions are somewhat lower. Even though the variances of M-R/S are lower, the effect of the bias is so strong that the MSE of M-R/S is higher than the one of R/S (approximately 0.8 times higher). The difference is the lowest for the log-$t$ distribution. Interestingly, the bias increases the MSE so much that the MSE increases with a time series length for M-R/S. For R/S, the MSE standardly decreases with the number of observations.

\subsection{AR(1)}

For short-range dependent AR(1), the results are quite as expected but not as strong. Even though R/S overestimates $H$ for a strong majority of distributions and lengths of the series, the maximum bias is 0.0309 for Laplace distribution of length $2^9$. The bias decreases with the number of observations for all but log-$t$ distribution. This is caused by the fact that the estimator is unbiased for log-$t$ distribution for small samples. For M-R/S, the estimates are less biased than the ones of R/S for majority of distributions. However, M-R/S is more biased for log-$t$ and log-Laplace distributions for all sample lengths. Moreover, the bias of M-R/S does not decrease with sample length in all cases. Similarly to the $i.i.d.$ case, the estimates are pushed downwards for log-$t$ and log-Laplace distributions. While this is true for both R/S and M-R/S, the effect is different. For R/S, the bias is pushed from the positive numbers closer to zero. For M-R/S, the bias is negative and pushed further away from zero. 

Even though MSE is dominated by variance for M-R/S, the effect of the bias is stronger for R/S. Nevertheless, the overall qualitative results remain the same for both the variance and MSE of the estimators. The variance and MSE of M-R/S are lower for shorter series. However, the difference becomes negligible for longer series. Note that again the variance and MSE are lower for log-$t$ and log-Laplace distributions.

\section{Conclusions}

In this paper, we analyzed the robustness of rescaled range analysis and modified rescaled range analysis to different distributional and memory properties. The bias, variance and mean squared errors of the estimators have been studied for independent, short-range dependent and long-range dependent processes with innovations from 8 different distributions. The simulations study uncovered mainly four interesting implications. 

First, extreme excess kurtosis combined with high skewness cause the estimates to be downward biased. Second, M-R/S in its simplest way is not able to distinguish between short and long memory efficiently. This is a very crucial result. It implies that M-R/S should not be used for Hurst exponent estimation in presence of short-term memory in the underlying process. This result is in hand with Teverosky \textit{et al.} \cite{Teverosky1999}, who showed that M-R/S is not suitable for adjusting the rescaled ranges. However, this should not disqualify the method altogether.  It shows that the method needs to be used properly and in conjunction with other methods. One of the examples is using M-R/S with bootstrapped confidence intervals as has been already done several times \citep{Kristoufek2010a,Onali2011}. This way, both types of dependence can be efficiently distinguished. Third, R/S is biased by the short-range dependence in the series but in a lesser way than standardly claimed. On contrary, R/S is less biased than M-R/S for some distributions. And fourth, apart from very extreme cases of high leptokurtosis and skewness, both methods are quite robust to different distributional properties.     

\section*{Acknowledgements}
The author would like to thank the anonymous referees for valuable comments and suggestions which helped to improve the paper significantly. The support from the Grant Agency of Charles University (GAUK) under project $118310$, Grant Agency of the Czech Republic (GACR) under project P402/11/0948, Ministry of Education MSMT 0021620841 and project SVV 261 501 are gratefully acknowledged.

\bibliography{MRS}
\bibliographystyle{chicago}

\newpage

\begin{table}[c]
\centering
\caption{Results of Monte Carlo simulations for R/S -- H=0.5}
\label{tab1}
\footnotesize
\begin{tabular}{|c|c|cccccc|}
\toprule \toprule
&&512&1024&2048&4096&8192&16384\\
\midrule \midrule
bias&normal&0.0034&0.0007&0.0038&0.0011&0.0005&0.0015\\
&log-normal&0.0003&-0.0026&0.0014&0.0038&0.0042&0.0053\\
&log-$t$&-0.0262&-0.0219&-0.0208&-0.0209&-0.0206&-0.0200\\
&Cauchy&-0.0088&-0.0090&-0.0049&0.0012&0.0028&0.0052\\
&Laplace&0.0043&0.0019&0.0069&0.0050&0.0060&0.0050\\
&log-Laplace&-0.0154&-0.0118&-0.0110&-0.0099&-0.0065&-0.0066\\
&gamma&0.0000&0.0028&0.0055&0.0017&0.0009&0.0007\\
&inv-gamma&-0.0002&0.0023&0.0019&0.0041&0.0047&0.0030\\
\midrule
variance&normal&0.0070&0.0036&0.0024&0.0015&0.0010&0.0008\\
&log-normal&0.0062&0.0036&0.0022&0.0016&0.0011&0.0007\\
&log-$t$&0.0033&0.0019&0.0011&0.0006&0.0004&0.0002\\
&Cauchy&0.0044&0.0027&0.0018&0.0013&0.0009&0.0006\\
&Laplace&0.0063&0.0038&0.0023&0.0016&0.0010&0.0007\\
&log-Laplace&0.0044&0.0027&0.0017&0.0011&0.0008&0.0006\\
&gamma&0.0058&0.0036&0.0023&0.0014&0.0011&0.0008\\
&inv-gamma&0.0058&0.0035&0.0022&0.0014&0.0011&0.0008\\
\midrule
MSE&normal&0.0070&0.0036&0.0024&0.0015&0.0010&0.0008\\
&log-normal&0.0061&0.0036&0.0022&0.0016&0.0011&0.0007\\
&log-$t$&0.0040&0.0024&0.0015&0.0011&0.0008&0.0006\\
&Cauchy&0.0045&0.0028&0.0018&0.0013&0.0010&0.0007\\
&Laplace&0.0064&0.0038&0.0023&0.0017&0.0011&0.0008\\
&log-Laplace&0.0046&0.0028&0.0018&0.0012&0.0008&0.0006\\
&gamma&0.0058&0.0036&0.0024&0.0014&0.0011&0.0008\\
&inv-gamma&0.0058&0.0035&0.0022&0.0014&0.0011&0.0008\\
\midrule
\end{tabular}
\end{table}

\begin{table}[c]
\centering
\caption{Results of Monte Carlo simulations for M-R/S -- H=0.5}
\label{tab2}
\footnotesize
\begin{tabular}{|c|c|cccccc|}
\toprule \toprule
&&512&1024&2048&4096&8192&16384\\
\midrule \midrule
bias&normal&-0.0071&-0.0068&-0.0017&-0.0032&-0.0030&-0.0015\\
&log-normal&-0.0085&-0.0084&-0.0030&0.0005&0.0016&0.0031\\
&log-$t$&-0.0336&-0.0266&-0.0241&-0.0234&-0.0225&-0.0215\\
&Cauchy&-0.0183&-0.0160&-0.0100&-0.0029&-0.0005&0.0022\\
&Laplace&-0.0070&-0.0059&0.0010&0.0006&0.0024&0.0020\\
&log-Laplace&-0.0235&-0.0171&-0.0149&-0.0129&-0.0088&-0.0084\\
&gamma&-0.0105&-0.0045&0.0001&-0.0025&-0.0025&-0.0022\\
&inv-gamma&-0.0098&-0.0046&-0.0028&0.0003&0.0017&0.0005\\
\midrule
variance&normal&0.0067&0.0035&0.0024&0.0015&0.0010&0.0008\\
&log-normal&0.0060&0.0035&0.0021&0.0015&0.0011&0.0007\\
&log-$t$&0.0034&0.0018&0.0011&0.0006&0.0004&0.0002\\
&Cauchy&0.0043&0.0027&0.0018&0.0013&0.0009&0.0006\\
&Laplace&0.0064&0.0038&0.0023&0.0016&0.0010&0.0007\\
&log-Laplace&0.0043&0.0027&0.0017&0.0011&0.0008&0.0006\\
&gamma&0.0057&0.0035&0.0023&0.0014&0.0011&0.0008\\
&inv-gamma&0.0056&0.0034&0.0022&0.0014&0.0011&0.0007\\
\midrule
MSE&normal&0.0068&0.0036&0.0024&0.0015&0.0010&0.0008\\
&log-normal&0.0061&0.0036&0.0021&0.0015&0.0011&0.0007\\
&log-$t$&0.0046&0.0025&0.0016&0.0012&0.0009&0.0007\\
&Cauchy&0.0047&0.0030&0.0019&0.0013&0.0009&0.0006\\
&Laplace&0.0064&0.0038&0.0023&0.0016&0.0010&0.0007\\
&log-Laplace&0.0049&0.0030&0.0019&0.0013&0.0008&0.0006\\
&gamma&0.0058&0.0035&0.0023&0.0014&0.0011&0.0008\\
&inv-gamma&0.0057&0.0034&0.0022&0.0014&0.0011&0.0007\\
\midrule
\end{tabular}
\end{table}

\begin{table}[c]
\centering
\caption{Results of Monte Carlo simulations for R/S -- H=0.75}
\label{tab3}
\footnotesize
\begin{tabular}{|c|c|cccccc|}
\toprule \toprule
&&512&1024&2048&4096&8192&16384\\
\midrule \midrule
bias&normal&-0.0010&-0.0031&-0.0124&-0.0278&-0.0439&-0.0600\\
&log-normal&-0.0021&-0.0005&-0.0121&-0.0266&-0.0433&-0.0618\\
&log-$t$&-0.0170&-0.0287&-0.0384&-0.0603&-0.0786&-0.0969\\
&Cauchy&-0.0041&-0.0022&-0.0189&-0.0344&-0.0512&-0.0656\\
&Laplace&0.0016&0.0013&-0.0127&-0.0262&-0.0437&-0.0608\\
&log-Laplace&-0.0041&-0.0119&-0.0208&-0.0421&-0.0569&-0.0745\\
&gamma&0.0029&-0.0001&-0.0162&-0.0282&-0.0449&-0.0596\\
&inv-gamma&-0.0050&-0.0045&-0.0102&-0.0272&-0.0445&-0.0606\\
\midrule
variance&normal&0.0109&0.0057&0.0032&0.0021&0.0013&0.0008\\
&log-normal&0.0096&0.0054&0.0034&0.0020&0.0013&0.0008\\
&log-$t$&0.0089&0.0041&0.0022&0.0011&0.0007&0.0004\\
&Cauchy&0.0079&0.0039&0.0025&0.0015&0.0011&0.0008\\
&Laplace&0.0102&0.0057&0.0033&0.0022&0.0013&0.0009\\
&log-Laplace&0.0083&0.0046&0.0027&0.0015&0.0009&0.0006\\
&gamma&0.0093&0.0054&0.0032&0.0022&0.0013&0.0009\\
&inv-gamma&0.0099&0.0052&0.0034&0.0021&0.0014&0.0008\\
\midrule
MSE&normal&0.0109&0.0057&0.0034&0.0029&0.0032&0.0044\\
&log-normal&0.0096&0.0054&0.0035&0.0027&0.0032&0.0046\\
&log-$t$&0.0092&0.0049&0.0037&0.0048&0.0068&0.0097\\
&Cauchy&0.0079&0.0039&0.0028&0.0027&0.0037&0.0051\\
&Laplace&0.0102&0.0057&0.0035&0.0029&0.0032&0.0046\\
&log-Laplace&0.0083&0.0047&0.0031&0.0033&0.0041&0.0062\\
&gamma&0.0093&0.0054&0.0035&0.0029&0.0033&0.0044\\
&inv-gamma&0.0099&0.0052&0.0035&0.0028&0.0033&0.0045\\
\midrule
\end{tabular}
\end{table}

\begin{table}[c]
\centering
\caption{Results of Monte Carlo simulations for M-R/S -- H=0.75}
\label{tab4}
\footnotesize
\begin{tabular}{|c|c|cccccc|}
\toprule \toprule
&&512&1024&2048&4096&8192&16384\\
\midrule \midrule
bias&normal&-0.1004&-0.1026&-0.1107&-0.1249&-0.1390&-0.1504\\
&log-normal&-0.0994&-0.1006&-0.1098&-0.1224&-0.1374&-0.1513\\
&log-$t$&-0.0896&-0.0947&-0.1072&-0.1258&-0.1446&-0.1605\\
&Cauchy&-0.0970&-0.0991&-0.1124&-0.1272&-0.1421&-0.1528\\
&Laplace&-0.0982&-0.0988&-0.1106&-0.1230&-0.1386&-0.1514\\
&log-Laplace&-0.0942&-0.1018&-0.1103&-0.1297&-0.1433&-0.1576\\
&gamma&-0.0963&-0.1000&-0.1144&-0.1251&-0.1394&-0.1500\\
&inv-gamma&-0.1020&-0.1034&-0.1089&-0.1239&-0.1386&-0.1506\\
\midrule
variance&normal&0.0074&0.0040&0.0024&0.0017&0.0011&0.0007\\
&log-normal&0.0067&0.0038&0.0025&0.0016&0.0011&0.0007\\
&log-$t$&0.0042&0.0021&0.0012&0.0006&0.0003&0.0002\\
&Cauchy&0.0051&0.0027&0.0019&0.0013&0.0009&0.0007\\
&Laplace&0.0069&0.0040&0.0025&0.0018&0.0011&0.0008\\
&log-Laplace&0.0052&0.0030&0.0018&0.0012&0.0007&0.0005\\
&gamma&0.0063&0.0037&0.0025&0.0017&0.0011&0.0008\\
&inv-gamma&0.0068&0.0037&0.0025&0.0017&0.0011&0.0007\\
\midrule
MSE&normal&0.0175&0.0145&0.0146&0.0173&0.0204&0.0234\\
&log-normal&0.0165&0.0139&0.0146&0.0166&0.0200&0.0236\\
&log-$t$&0.0123&0.0110&0.0127&0.0164&0.0212&0.0260\\
&Cauchy&0.0146&0.0125&0.0145&0.0174&0.0211&0.0240\\
&Laplace&0.0165&0.0138&0.0147&0.0169&0.0203&0.0237\\
&log-Laplace&0.0141&0.0134&0.0140&0.0180&0.0212&0.0254\\
&gamma&0.0156&0.0137&0.0155&0.0174&0.0206&0.0233\\
&inv-gamma&0.0172&0.0144&0.0144&0.0170&0.0204&0.0234\\
\midrule
\end{tabular}
\end{table}

\begin{table}[c]
\centering
\caption{Results of Monte Carlo simulations for R/S -- AR(1)}
\label{tab5}
\footnotesize
\begin{tabular}{|c|c|cccccc|}
\toprule \toprule
&&512&1024&2048&4096&8192&16384\\
\midrule \midrule
bias&normal&0.0288&0.0220&0.0216&0.0192&0.0173&0.0134\\
&log-normal&0.0253&0.0208&0.0180&0.0205&0.0194&0.0158\\
&log-$t$&0.0042&0.0022&-0.0026&-0.0048&-0.0055&-0.0063\\
&Cauchy&0.0158&0.0120&0.0139&0.0145&0.0165&0.0157\\
&Laplace&0.0309&0.0256&0.0232&0.0209&0.0186&0.0165\\
&log-Laplace&0.0123&0.0094&0.0072&0.0090&0.0058&0.0059\\
&gamma&0.0253&0.0220&0.0204&0.0177&0.0146&0.0149\\
&inv-gamma&0.0273&0.0226&0.0222&0.0186&0.0175&0.0164\\
\midrule
variance&normal&0.0066&0.0040&0.0023&0.0015&0.0010&0.0007\\
&log-normal&0.0061&0.0039&0.0024&0.0015&0.0011&0.0008\\
&log-$t$&0.0036&0.0018&0.0010&0.0005&0.0004&0.0002\\
&Cauchy&0.0045&0.0029&0.0018&0.0014&0.0009&0.0007\\
&Laplace&0.0064&0.0040&0.0025&0.0015&0.0011&0.0009\\
&log-Laplace&0.0044&0.0026&0.0017&0.0010&0.0007&0.0006\\
&gamma&0.0066&0.0037&0.0023&0.0016&0.0011&0.0008\\
&inv-gamma&0.0060&0.0038&0.0024&0.0017&0.0010&0.0008\\
\midrule
MSE&normal&0.0074&0.0045&0.0028&0.0019&0.0013&0.0009\\
&log-normal&0.0067&0.0043&0.0027&0.0019&0.0014&0.0010\\
&log-$t$&0.0036&0.0018&0.0010&0.0005&0.0004&0.0003\\
&Cauchy&0.0047&0.0030&0.0020&0.0016&0.0011&0.0009\\
&Laplace&0.0073&0.0046&0.0030&0.0019&0.0014&0.0011\\
&log-Laplace&0.0045&0.0027&0.0018&0.0011&0.0007&0.0006\\
&gamma&0.0072&0.0042&0.0027&0.0019&0.0013&0.0010\\
&inv-gamma&0.0068&0.0043&0.0028&0.0020&0.0013&0.0010\\
\midrule
\end{tabular}
\end{table}

\begin{table}[c]
\centering
\caption{Results of Monte Carlo simulations for M-R/S -- AR(1)}
\label{tab6}
\footnotesize
\begin{tabular}{|c|c|cccccc|}
\toprule \toprule
&&512&1024&2048&4096&8192&16384\\
\midrule \midrule
bias&normal&-0.0019&-0.0052&-0.0019&-0.0023&-0.0022&-0.0044\\
&log-normal&-0.0055&-0.0066&-0.0061&-0.0015&-0.0004&-0.0019\\
&log-$t$&-0.0246&-0.0244&-0.0256&-0.0255&-0.0246&-0.0235\\
&Cauchy&-0.0143&-0.0139&-0.0095&-0.0071&-0.0030&-0.0022\\
&Laplace&0.0006&-0.0011&-0.0009&-0.0011&-0.0009&-0.0014\\
&log-Laplace&-0.0182&-0.0170&-0.0170&-0.0124&-0.0135&-0.0115\\
&gamma&-0.0056&-0.0050&-0.0037&-0.0044&-0.0049&-0.0029\\
&inv-gamma&-0.0034&-0.0049&-0.0016&-0.0032&-0.0020&-0.0016\\
\midrule
variance&normal&0.0060&0.0037&0.0022&0.0015&0.0010&0.0007\\
&log-normal&0.0055&0.0036&0.0023&0.0014&0.0010&0.0008\\
&log-$t$&0.0033&0.0016&0.0009&0.0005&0.0004&0.0002\\
&Cauchy&0.0041&0.0027&0.0017&0.0013&0.0008&0.0007\\
&Laplace&0.0059&0.0037&0.0023&0.0014&0.0010&0.0008\\
&log-Laplace&0.0040&0.0024&0.0016&0.0010&0.0007&0.0006\\
&gamma&0.0060&0.0035&0.0022&0.0016&0.0011&0.0007\\
&inv-gamma&0.0054&0.0035&0.0022&0.0016&0.0010&0.0008\\
\midrule
MSE&normal&0.0060&0.0037&0.0022&0.0015&0.0010&0.0007\\
&log-normal&0.0055&0.0037&0.0023&0.0014&0.0010&0.0008\\
&log-$t$&0.0039&0.0022&0.0016&0.0012&0.0010&0.0008\\
&Cauchy&0.0043&0.0029&0.0018&0.0014&0.0008&0.0007\\
&Laplace&0.0059&0.0037&0.0023&0.0014&0.0010&0.0008\\
&log-Laplace&0.0043&0.0027&0.0019&0.0011&0.0009&0.0007\\
&gamma&0.0060&0.0035&0.0022&0.0016&0.0011&0.0007\\
&inv-gamma&0.0054&0.0035&0.0022&0.0016&0.0010&0.0008\\
\midrule
\end{tabular}
\end{table}

\end{document}